\documentstyle[12pt]{article}
\begin{document}
\newcommand{\newc}{\newcommand}
\newc{\gev}{\,GeV}
\newc{\rp}{$R_p$}
\newc{\rpv}{$\not\!\!R_p$}
\newc{\rpvm}{{\not\!\! R_p}}
\newc{\rpvs}{{\not R_p}}
\newc{\ra}{\rightarrow}
\newc{\lra}{\leftrightarrow}
\newc{\lsim}{\buildrel{<}\over{\sim}}
\newc{\gsim}{\buildrel{>}\over{\sim}}
\newc{\esim}{\buildrel{\sim}\over{-}}
\newc{\lam}{\lambda}
\newc{\lsp}{{\tilde\Lambda}}
\newc{\oc}{{\cal {O}}}
\newc{\msq}{m_{\tilde q}}
\newc{\mpl}{M_{Pl}}
\newc{\mw}{M_W}
\newc{\pho}{{\tilde\gamma}}
\newc{\half}{\frac{1}{2}}
\newc{\third}{\frac{1}{3}}
\newc{\fourth}{\frac{1}{4}}
\newc{\beq}{\begin{equation}}
\newc{\eeq}{\end{equation}}
\newc{\barr}{\begin{eqnarray}}
\newc{\earr}{\end{eqnarray}}
\newc{\ptmis}{\not\!\!{p_T}}
\newc{\mc}{\multicolumn}
\newc{\ol}{\overline}
\newc{\ecal}{{\cal E}}
\newc{\Xs}{cross section}
\newc{\delb}{\Delta B\not=0}
\newc{\dell}{\Delta L\not=0}
\newc{\delbi}{\Delta B_i\not=0}
\newc{\delli}{\Delta L_i\not=0}
\title{Ansatz for Quark, Charged Lepton, and Neutrino Masses in SUSY GUTS}
\author{H.~Dreiner$^1$, G. K. Leontaris$^2$ and N. D. Tracas$^3$}
\date{{\small $^1$Department of Physics, University of Oxford,\\ 1 Keble Rd,
Oxford  OX1 3NP, UK\\
$^2$ Physics Dept, Theory Div., University of Ioannina, Ioannina, Greece \\
$^3$ National Technical University, Athens, Greece}}
\maketitle

\begin{abstract}
\noindent We extend a fermion mass matrix Ansatz by Giuduce to include neutrino
masses. The previous predictions are maintained. With two additional
parameters,
a large Majorana neutrino mass and a hierarchy factor, we have seven {\it
further} low energy predictions: the masses of the neutrinos, the mixing angles
and the phase in the leptonic sector. We choose a reasonable hierarchy of
Majorana masses and fit the overall mass scale according to a solution of the
solar neutrino problem via the MSW mechanism, which is in agreement with
the $^{37}Cl$, Kamiokande, and GALLEX data. We then also obtain a
cosmologically interesting tau-neutrino mass.
\end{abstract}

One of the fundamental problems of particle physics is to understand the
observed fermion mass spectrum. In the Standard Model it is described by
thirteen parameters: nine masses, three quark mixing angles and a $CP$
violating
phase. Ultimately one hopes for a solution in terms of a more fundamental
model, {\it e.g.} string theory, where the structure of the mass matrices is
determined by a set of symmetries. Meanwhile any reduction in the set of
required parameters could be a guide in finding the appropriate symmetries and
also lead to a prediction of yet undetermined fermion masses in terms of
experimentally known masses and angles.

There has been much fruitful work on this problem \cite{fritzsch,dhr,giudice}.
Recently, motivated by the observed merging of the Standard Model gauge
coupling constants in supersymmetric grand unified theories (SUSY GUTs)
\cite{lang}, Dimopoulos, Hall and Raby (DHR) \cite{dhr} and Giudice
\cite{giudice} have considered simple fermion mass matrices at the GUT scale
and derived the resulting spectrum below the electroweak scale using the
renormalization group equations (RGEs) for the case of minimal supersymmetry.
In the DHR ansatz the fourteen parameters (including $\tan\beta$, the ratio of
Higgs vevs in supersymmetry) are given in terms of eight input parameters;
hence six predictions. In the modified ansatz studied by Giudice more structure
is imposed on the mass matrices (by hand) leading to one input parameter less
and thus one more prediction. Both are in good agreement with the known
experimental values \cite{dhr,barger,giudice}. In a large class of mass
matrices
these ans\"atze are the only consistent solutions \cite{dickgraham}.

In the work of DHR and of Giudice neutrino mass matrices were not considered.
Most GUT models, {\it e.g.} $SO(10)$, $SU(4)\otimes SU(2)^2$, $SU(5)\otimes
U(1)$, and $E_6$, predict the existence of right-handed neutrinos leading to
Dirac-neutrino masses at the GUT scale. In this letter we extend the ansatz by
Giudice to include neutrino masses. This modifies the analysis in the following
points:

(i) New Higgs representations should provide heavy Majorana masses to the
right-handed neutrinos, in order to realize the see-saw mechanism
\cite{ramond}.

(ii) The leptonic mixing angles are completely determined by the charged-lepton
and up-quark masses. With one additional parameter we obtain the neutrino mass
spectrum and the mixing angles in the lepton sector. We thus have seven further
low-energy predictions. We obtain a possible solution to the solar neutrino
problem via the MSW mechanism \cite{msw} in agreement with the combined data of
the $^{37}Cl$, Kamiokande, and GALLEX experiments \cite{gallex} and the
tau-neutrino mass is cosmologically relevant.

\bigskip

A realistic fit to the desired structure of the mass matrices in SUSY GUT
models
requires a rather complicated Higgs sector. In a more fundamental theory the
structure of the matrices might arise more naturally. However, as an example we
consider here the relevant couplings in an $SO(10)$ SUSY GUT, assuming that
additional discrete symmetries, as in Harvey {\it et al.} of
Ref.\cite{fritzsch}, prevent the appearance of specific entires in the mass
matrices..

Within an $SO(10)$ SUSY GUT the down-like quark and charged lepton masses
arise from the following terms
\beq
{\hat f}( {\underline{16}} _1 \cdot{\underline{16}} _2\cdot
{\underline{10}}_{<{\bar 5}>}) + (2{\hat d}) ({\underline{16}} _2
\cdot{\underline{16}} _3 \cdot {\underline{10}}_{<{\bar 5}>})
\\
+ {\hat d} ({\underline{16}} _2\cdot{\underline{16}}
_2\cdot{\underline{126}}_{<{\bar {45}}>}) + {\hat c} (
{\underline{16}}_3\cdot {\underline{16}}_3
\cdot {\underline{10}}_{<{\bar 5}>}) ,
\label{eq:sotendown}
\eeq
where ${\underline {16}}_i$ are the fermion families. The simple relation
between the second and third coefficient is imposed by hand. The
${\underline{126}}$ Higgs representation has a non-zero vev along the
$<\overline{45}>$ direction of the $SU(5)$. The structure of this vev results
in a relativ factor $(-3)$ in the (2,2)-entry of the mass matrices $M_d$
and $M_e$ below \cite{frampton,georgijarlskog} and leads to the successful
Georgi-Jarlskog relations $m_b\simeq m_\tau$, $m_s \simeq m_\mu/3$, and
$m_d\simeq 3m_e$ at the GUT-scale \cite{georgijarlskog}. The up-quark and the
Dirac-neutrino masses arise from
\beq
{\hat b}({\underline{16}}_1\cdot{\underline{16}}_3 \cdot
{\underline{10}}_{<5>}+ {\underline{16}}_2
\cdot{\underline{16}}_2\cdot{\underline{10}}_{<5>}) + {\hat a}\,(
 {\underline{16}}_3\cdot
{\underline{16}}_3\cdot{\underline{10}}_{<5>}).
\label{eq:sotenup}
\eeq
and the heavy Majorana-neutrino mass matrix from
\beq
\lam_{\nu_i} {\underline{16}}_i \cdot {\underline{16}}_i
<{\underline{126}}>_{<1>} \ra M_i\nu_i^c\nu_i^c.
\label{eq:sotenmaj}
\eeq
where we have assumed a diagonal Majorana mass matrix, $diag(M_1,M_2,M_3)$. In
the following we shall make the simple hierarchical ansatz $M=M_3=k
M_2=k^2M_1$,
$k=10$, motivated by the known fermion mass hierarchies. $\lam$ is a Yukawa
coupling and $i=1,2,3$ is a generation index.

We obtain the Giudice mass matrices \cite{giudice} augmented by a simple
ansatz for the neutrino masses
\barr
M_u&=&\left(\begin{array}{ccc} 0&0&b\\0&b&0\\b&0&a\end{array} \right), \qquad
\quad
M_{\nu\nu^c}=\left(\begin{array}{ccc} 0&0&b\\0&b&0\\b&0&a\end{array} \right),
\label{eq:upmat}\\ &&  \nonumber \\
M_d&=& \left(\begin{array}{ccc} 0&f e^{i\phi}&0\\f e^{-i\phi}&d&2d\\0&2d&c
\end{array} \right), \quad
M_e = \left(\begin{array}{ccc} 0&f e^{i\phi}&0\\f e^{-i\phi}&-3d&2d\\0&2d&c
\end{array} \right), \label{eq:elecmat}\\ && \nonumber \\
&& \quad M_{\nu^c\nu^c}= M\, diag(k^{-2},k^{-1},1). \label{eq:majmat}
\earr
The entries in the mass matrices are given by the corresponding Yukawa
coupling of Eq.(\ref{eq:sotendown}-\ref{eq:sotenmaj}) multiplied with the
appropriate vev. In general these entries are complex. However, the fields can
be redefined such that all parameters are real and we only have the shown
phases \cite{giudice}. We have included a single phase in the charged lepton
mass matrix, since the neutrinos are no longer degenerate. For simplicity we
have chosen this phase to be identical with the phase in $M_d$. \footnote{In
general the leptonic mixing matrix will include three phases.} We thus have
eight parameters at the GUT scale (including $k$) in order to provide a total
of 20 parameters at low energies: six quark masses; three charged lepton
masses; three light neutrino masses; six mixing angles, three for the
Cabbibo-Kobayashi-Maskawa mixing matrix and three for the corresponding
leptonic mixing matrix; and two phases, one for each mixing matrix. Hence, we
have seven predictions for the Standard Model parameters and seven predictions
for the extension of the Standard Model due to the massive neutrinos.

Using the results of Ref.\cite{falck} we obtain the renormalization group
equations for the Yukawa couplings at the one-loop level
\barr
16\pi^2 \frac{d}{dt} \lam_U&=& \left( I\cdot Tr [3 \lam_U\lam_U^\dagger ]  +
3 \lam_U \lam_U^\dagger +\lam_D \lam_D^\dagger
-I\cdot G_U\right) \lam_U, \label{eq:rge1}
\\
16\pi^2 \frac{d}{dt} \lam_N&=& \left( I\cdot Tr [
\lam_U \lam_U^\dagger ]  + \lam_E \lam_E^\dagger -I
\cdot G_N\right) \lam_N, \label{eq:rge2}
\\
16\pi^2 \frac{d}{dt} \lam_D&=& \left( I\cdot Tr [\lam_D\lam_D^\dagger + \lam_E
\lam_E^\dagger ]  + 3 \lam_D \lam_D^\dagger +\lam_U \lam_U^\dagger -I \cdot
G_D\right) \lam_D, \label{eq:rge3}
\\
16\pi^2 \frac{d}{dt} \lam_E&=& \left( I\cdot Tr [ \lam_E\lam_E^\dagger + \lam_D
\lam_D^\dagger ]  + 3 \lam_E \lam_E^\dagger -I \cdot G_E\right) \lam_E,
\label{eq:rge4}
\earr
where $\lam_\alpha$, $\alpha=U,N,D,E$, represent the $3$x$3$ Yukawa matrices
which are defined in terms of the mass matrices given in
Eq.(\ref{eq:upmat}-\ref{eq:majmat}), and $I$ is the $3$x$3$ identity matrix.
We have neglected one-loop corrections proportional to $\lam_N^2$.
$t\equiv\ln(\mu/\mu_0)$, $\mu$ is the scale at which the couplings are to be
determined and $\mu_0$ is the reference scale, in our case the GUT scale. The
gauge contributions are given by
\barr
G_\alpha&=&\sum_{i=1}^3 c_\alpha^i g_i^2(t),\\
g_i^2(t)&=&\frac{g_i^2(t_0)}{1- \frac{b_i}{8\pi^2} g_i^2(t_0)(t-t_0)}.
\earr
The $g_i$ are the three gauge coupling constants of the Standard Model and
$b_i$
are the corresponding beta functions in minimal supersymmetry. The coefficients
$c_\alpha^i$ are given by
\barr
\{c_U^i \}_{i=1,2,3} &=& \left\{ \frac{13}{15},3,\frac{16}{3} \right\}, \qquad
\{c_D^i \}_{i=1,2,3} = \left\{\frac{7}{15},3,\frac{16}{3} \right\}, \\
\{c_E^i \}_{i=1,2,3} &=& \left\{ \frac{9}{5},3,0\right\}, \qquad \quad
\{c_N^i \}_{i=1,2,3} = \left\{ \frac{3}{5},3,0\right\}.
\earr

We find it convenient to redefine the quark and lepton fields such that
$\lam_U$ {\it and} $\lam_N$ are diagonal
\barr
\lam_U\ra {\tilde\lam}_U&=& K^\dagger\lam_UK,\qquad {\lam}_N\ra
{\tilde\lam}_N= K^\dagger\lam_NK, \nonumber \\
\lam_D\ra {\tilde\lam}_D&=& K^\dagger\lam_DK,\qquad
\lam_E\ra {\tilde\lam}_E= K^\dagger\lam_EK.  \label{eq:redef}
\earr
The diagonalizing matrix is given by \cite{giudice}
\beq
K=\left( \begin{array}{ccc} \cos\theta & 0& \sin\theta \\ 0&1&0\\
-\sin\theta&0&\cos\theta \end{array} \right), \quad \tan 2\theta=\frac{2b}{a}.
\eeq
For the up-quarks and the corresponding Dirac-neutrinos we obtain at the GUT
scale
\barr
m_u&=&m_{\nu_{D1}}= m_0 \tan^2\theta, \quad m_c=m_{\nu_{D2}}= m_0 \tan\theta,
 \nonumber \\
m_t&=&m_{\nu_{D3}} =m_0, \quad m_0= \frac{a}{1-\tan^2\theta}.
\label{eq:upmass}
\earr
We apply the field redefinitons (\ref{eq:redef}) to the differential equations
(\ref{eq:rge1}-\ref{eq:rge4}) and within the parenthesis on the right hand side
only retain the dominant Yukawa coupling ${\tilde\lam}_{U_{33}}^2$
\barr
16\pi^2 \frac{d}{dt} {\tilde{\lam}}_U&=& (
{\tilde{\lam}}^2_{U_{33}}
\left( \begin{array}{ccc} 3&& \\ &3&
\\&&6
\end{array}
\right)
-G_U(t) \,I  ) {\tilde{\lam}}_U , \label{eq:rges1}
\\
16\pi^2 \frac{d}{dt} {\tilde{\lam}}_N&=&
 (
 {\tilde{\lam}}_{U_{33}}^2 -G_N(t)) \,I )
) {\tilde{\lam}}_N, \label{eq:rges2}
\\
16\pi^2
\frac{d}{dt}{\tilde{\lam}}_D&=& (  {\tilde{\lam}}^2_{U_{33}} \left(
\begin{array}{ccc} 0&&\\&0&\\&&1\end{array}\right) -G_D(t) \,I)
{\tilde{\lam}}_D, \label{eq:rges3}
\\
16\pi^2 \frac{d}{dt}{\tilde{\lam}}_E&=&
 -G_E(t) \,I \,{\tilde{\lam}}_E.
\label{eq:rges4}
\earr
The equations for ${\tilde\lam}_U, {\tilde\lam}_D$, and ${\tilde \lam}_E$ are
thus unaffected by the neutrinos. We have solved this (approximate) system by
first obtaining the exact solution for ${\tilde \lam}_{U_{33}}$
\cite{lopez,giudice}
\beq
{\tilde\lam}_{U_{33}}(t) = {\tilde\lam}_{U_{33}} (t_0)
\zeta^6 \gamma_{U} (t)
\eeq
where
\barr
\gamma_\alpha(t)&=& \exp({-\int G_\alpha(t) \,dt/(16\pi^2)})\\
&=& \prod_{j=1}^3 \left( \frac{\alpha_{j,0}}{\alpha_j}\right)^{c_\alpha^j/2b_j}
\\
&=& \prod_{j=1}^3 \left(1- \frac{b_{j,0}\alpha_{j,0}(t-t_0)}{2\pi}
\right)^{c_\alpha^j/2b_j},
\\
\zeta&=& \exp({\frac{1}{16\pi^2}\int_{t_0}^t \lam^2_{U_{33}}dt}) \\
&=&\left( 1-\frac{\kappa}{8\pi^2}\lam_\alpha(t_0)
\int_{t_0}^{t} \gamma_\alpha^2(t)\,dt \right)^{-1/(2\kappa)}.
\earr
We then use this solution for ${\tilde\lam}_{U_{33}}$ to solve the
equations for the other couplings. We reproduce the results of
\cite{giudice}, including
\barr
m_b&=& \eta_b \frac {\gamma_D}{\gamma_E} \zeta    m_\tau, \\
m_t&=& \zeta^3 \frac{\eta_u}{\eta_c^2} \frac{m_c^2}{m_u}.
\label{eq:mt}
\earr
$\eta_q=m_q(1\gev)/m_q(m_t)$ for the light quarks and $\eta_q=m_q(m_q)/m_q(m_t)
$. for the $c$ and $b$ quark. In our analysis we have taken $\eta_u=\eta_d=
\eta_s =2$, $\eta_c=1.8$ and $\eta_b=1.3$. $\gamma_D/\gamma_E=2.1$ and
$\gamma_U
/\gamma_N=2.4$. In order to produce the proper running bottom quark mass
\cite{narison,gasser}
\beq
m_b(m_b) = 4.25 \pm 0.1 \gev
\eeq
we require $\zeta=0.81\pm0.2$. Following \cite{giudice} we obtain the range
\beq
125\gev\leq m_t \leq 170\gev
\label{eq:heavymt}
\eeq
from bounds on $\tan\beta$ from electroweak breaking as well as on $m_u$
\cite{gasser}.

\bigskip

In order to determine the lepton mixing matrices we must find the matrix $V_l$
such that $(V_l^\dagger {\tilde\lam}^R_E ({\tilde\lam}^R_E)^\dagger  V_l)$ is
diagonal. We parametrize the mixing matrix as
\beq
V_l=\left( \begin{array}{ccc}
c_1c_3e^{i\phi}-s_1s_2s_3 & s_1c_3e^{i\phi}+c_1s_2s_3& -c_2s_3 \\
-s_1c_2 &c_1c_2& s_2 \\
c_1s_3e^{i\phi}+s_1s_2c_3&s_1s_3e^{i\phi}-c_1s_2c_3& c_2c_3
\end{array} \right),
\eeq
where $c_1= \cos\theta_1$, $s_1=\sin\theta_1$, {\it etc}. Then we find
\barr
V_{\nu_\mu e}&\simeq&|s_1|=\sqrt{y} (1-\half y)\simeq 6.9\cdot 10^{-2} ,
\label{eq:lepmix1}\\
V_{\nu_\mu\tau}&=&|s_2|= \frac{2}{3} x(1-y-\frac{13}{9}  x)\simeq (3.95
\pm0.01)
10^{-2},
\label{eq:lepmix2}\\
V_{\nu_e\tau}&=&|s_3|= \frac{m_c}{\eta_c m_t}\simeq (4.9\pm0.2) 10^{-3}
\left( \frac{145\gev} {m_t} \right) .
\label{eq:lepmix3}
\earr
where $x=m_\mu/m_\tau$, $y=m_e/m_\mu$ and we have used the values for the
charged lepton masses given by the particle data group \cite{partdata}. In the
last equation we have used the constraint on the top quark mass
(\ref{eq:heavymt}) and $m_c=(1.27\pm0.05)\gev$. Thus we predict the ``leptonic
Cabbibo angle" ($\theta_1$) to be substantially smaller than the corresponding
quark angle. This is different from phenomenological ans\"atze for the angle in
Refs.\cite{bludman}, where the leptonic mixing angles are approximated by the
corresponding quark angles. It is directly due to the factor of $(-3)$ in the
down-quark mass matrix, which is {\it necessary} to obtain the proper mass
predictions. In ref. \cite{ellis} a similar value is obtained from a different
phenomenological ansatz.

The angle $\phi$ determines a $CP$ violating quantity $J$ discussed in
\cite{giudice}, which is experimentally only poorly determined. The
corresponding $J_l$ for the leptonic mixing matirx is smaller by about a factor
of 3, corresponding to $s_1^l\simeq\third s_1^q$.

{}From the equations (\ref{eq:lepmix1}-\ref{eq:lepmix3}) we obtain  the
following neutrino mixing angles which we choose to write in the form
$\sin^22\theta_{ex}$, motivated by the solutions to the solar neutrino problem
\barr
\sin^22\theta_{e\mu} &=& 1.9\cdot 10^{-2} \label{eq:emu} \\
\sin^22\theta_{\mu\tau} &=& (6.2\pm 0.1)10^{-3} \label{eq:mutau} \\
\sin^22\theta_{e\tau} &=& (9.6\pm 0.4 ) 10^{-5} \label{eq:etau}
\earr
and we have estimated the last angle by $\theta_3$ since this ansatz does not
determine the phase $\phi$. We thus obtain a mixing angle $\theta_{e\tau}$
which is smaller than the phenomenological ansatz in \cite{ellis} and is
presumably out of the reach of the NOMAD and CHORUS experiments.

\bigskip

Renormalizing $\lam_N$ down to $m_t$ and expressing the eigenvalues in terms
of the up-quark masses, the Dirac-neutrino masses are
\barr
m_{\nu_{D1}} &=& \frac{\gamma_N}{\gamma_U} \frac{1}{\eta_u\zeta^2}\, m_u,
\nonumber
\\
m_{\nu_{D2}} &=& \frac{\gamma_N}{\gamma_U} \frac{1}{\eta_c\zeta^2} \,m_c,
\nonumber
\\
m_{\nu_{D3}} &=& \frac{\gamma_N}{\gamma_U} \frac{1}{\zeta^5} \,m_t.
\label{eq:dnm}
\earr
In order to obtain realistic neutrino masses at the weak-scale we implement the
see-saw mechanism \cite{ramond} via the universal heavy Majorana mass matrix
(\ref{eq:majmat}). The matrix $K$, which diagonalizes the Dirac-neutrino mass
matrix produces off-diagonal elements in the heavy Majorana mass matrix
$M_{\nu^c\nu^c}$. However, to a good approximation the eigenvalues of
the complete $6$x$6$ neutrino mass matrix are unaffected by this rotation.
For each generation we thus have the neutrino mass matrix
\beq
\begin{tabular}{c|cc}
& $\nu_i$ & $\nu^c_i$ \\ \hline
$\nu_i$& 0 & $\half m_{\nu_{Di}}$ \\
$\nu^c_i$ & $\half m_{\nu_{Di}}$ &$M_i$
\end{tabular}
\eeq
Diagonalizing this matrix we obtain the light neutrino masses $m_{\nu_i} =
m_{\nu_{Di}}^2 /(4M_i)$, which we can express in terms of the up-quark
masses using Eqs.(\ref{eq:dnm})
\barr
m_{\nu_e}&=& \fourth (\frac{\gamma_N}{\gamma_U} \frac{1}{\eta_u\zeta^2})^2
\frac{k^2 m_u^2}{ M}= (2.4\pm0.2)  \frac{m_u^2}{M}, \label{eq:nue}\\
m_{\nu_\mu}&=& \fourth (\frac{\gamma_N}{\gamma_U} \frac{1}{\eta_c\zeta^2})^2
\frac{km_c^2}{M} = (3.0 \pm 0.3) 10^{-1} \frac{m_c^2}{M},  \label{eq:numu}\\
m_{\nu_\tau}&=& \fourth (\frac{\gamma_N}{\gamma_U}
\frac{1}{\zeta^5})^2 \frac{m_t^2}{M} = (0.35\pm0.09) \frac{m_t^2}{M}
\label{eq:nutau} .
\earr
Recall that we have made the simple hierarchy ansatz $M=M_3=kM_2=k^2M_1$,
$k=10$, which also gives three heavy neutrinos which are inaccessible to
experiment. The ratio of the light neutrino masses is predicted to be
\beq
m_{\nu_e} : m_{\nu_\mu} : m_{\nu_\tau}= (10 m_u/\eta_u)^2 : (3.1 m_c/\eta_c)^2:
(m_t^2/\zeta^3),
\label{eq:neutratio}
\eeq
This is what one would expect in a general GUT see-saw mechanism \cite{ramond}.
It is similar to the phenomenological ansatz discussed in \cite{ellis}.
Due to our restrictive ansatz (\ref{eq:upmat}-\ref{eq:majmat}) we have the
additional constraint (\ref{eq:mt}) and $m_t$ is {\it not} a free parameter. As
seen above, the mixing angles in the leptonic sector are determined through
known quantities. So with just one additional parameter, $M$, and a hierarchy
,$k$, we fix the scale of the neutrino masses and have a completely determined
neutrino sector.

We chose to fix $M$ so as to obtain a solution to the solar neutrino problem
via the MSW mechanism \cite{msw}. We assume a two generation mixing solution
to this problem. From Fig.1 in the second reference \cite{gallex}, which
includes the $^{37}Cl$, the Kamiokande and the GALLEX data, we have two
possible solutions
\beq
\sin2\theta_{ex} = (0.39-2.2)10^{-2} ,\quad \Delta m^2= (0.29- 2.0) 10^{-5}
eV^2, \label{eq:sol1}
\eeq
and
\beq
\sin2\theta_{ex} = 0.2-0.8 ,\quad \Delta m^2= (0.31-9.1) 10^{-5} eV^2.
\label{eq:sol2}
\eeq
Comparing with Eqs(\ref{eq:emu}-\ref{eq:etau}), we have one possible solution
in the range of (\ref{eq:sol1}) for $\nu_e-\nu_\mu$ mixing. Using Fig. 1 of
ref\cite{gallex} this requires
\beq
m_{\nu_\mu} = (1.7-1.9) \, 10^{-3} eV.
\eeq
This fixes $M$ to be
\beq
M= (2.7\pm0.5)10^{11}\gev
\eeq
Using Eqs(\ref{eq:nue},\ref{eq:nutau}) we then predict the other neutrino
masses
\barr
m_{\nu_e} &=& (2.3\pm0.6)10^{-7} eV
\\
m_{\nu_\tau} &=& (27.3\pm 8.7)eV \left( \frac{m_t}{145\gev}\right)^2
\earr
The resulting tau-neutrino mass is in a cosmologically interesting range
\cite{kolbturner}. Using the formula $\Omega_\nu h^2=m_\nu/91.5\,eV$,
where $h=0.5-1$ is the Hubble parameter \cite{kolbturner}, for the relic
density of the neutrinos we obtain
\beq
\Omega_\nu h^2 = 0.27\pm0.07.
\eeq
This is in the range one would expect from the recent COBE data.

\bigskip

In conclusion we have expanded an ansatz bei Giudice for the fermion mass
matrices at the GUT scale to include massive neutrinos, since these are
predited by most GUTs. Giudice's predictions for the Standard Model parameters
are retained. Without any additional parameters we predict the neutrino
mixing angles completely in terms of known fermion masses. In addition with one
parameter and a simple hierarchy we predict a neutrino mass spectrum, which
gives a possible solution to the solar neutrino problem via the MSW mechanism
and gives a significant contribution to the hot dark matter in agreement with
recent interpretations of the COBE data.

\bigskip
We wish to thank the CERN theory division for its hospitality and financial
support, while this work was done. Two of us (H.D. and G.L.) also acknowledge
partial support by the EC-grant SCI-0221-C(TT).

\bigskip

\noindent {\bf Note Added:} While this letter was being written we received a
copy of Ref.\cite{dhr2} in which DHR extend their original ansatz to include
neutrino masses as well. They have one fewer parameter in the neutrino sector
and obtain a lighter tau-neutrino mass. Since this is a different mass matrix
ansatz our work is complimentary to their's.

We also just received a copy of \cite{uri} where among other topics the
constraint on the Giudice ansatz due to CP-asymmetries in B-meson decays
is considered. This leads to stronger bounds on the quark mixing angles and
indirectly to a stronger bound on $m_t$ than our Eq.(\ref{eq:heavymt}):
$125\leq m_t\leq 155\gev$ \cite{uri}.

\end{document}